\documentclass[twocolumn,showpacs,showkeywords,prb]{revtex4}
\usepackage{graphicx}
\usepackage{dcolumn}
\usepackage{amsmath}
\usepackage{latexsym}

\begin{document}
\title
{Noncommutative quantum mechanics of a harmonic oscillator
under linearized\\ gravitational waves}
\author{Anirban Saha}
\email{anirban@iucaa.ernet.in }
%\affiliation{Department of Physics and Astrophysics, 
%West Bengal State University,\\
%Barasat, North 24 Paraganas,\\ West Bengal, India\\ }
\altaffiliation{Visiting Associate in Inter University Centre for Astronomy
and Astrophysics, Pune, India}
\author{Sunandan Gangopadhyay}
\email{sunandan@bose.res.in, sunandan.gangopadhyay@gmail.com }
%\email{sunandan.gangopadhyay@gmail.com}
\altaffiliation{Visiting Associate in Satyendra Nath Bose National Centre for
Basic Sciences, Kolkata, India}
\affiliation{Department of Physics and Astrophysics, 
West Bengal State University,\\
Barasat, North 24 Paraganas,\\ West Bengal, India\\ }

\author{Swarup Saha}
\affiliation{Debipur, Duttapukur 743248, India\\ }
%\date{\today}

%\author{ Anirban Saha\\ Department of Physics, Sovarani Memorial College\\ Jagatballavpur, Howrah - 711 408, West Bengal, India} 
%\cos \Lambda t \left(1 + \frac{f ^{2}\left(t \right)}{2}\right) 
\begin{abstract}
{\noindent We consider the quantum dynamics of a harmonic oscillator in noncommutative space under the influence of linearized gravitational waves (GW) in the long wave-length and low-velocity limit. Following the prescription in \cite{ncgw1} we quantize the system. The Hamiltonian of the system is solved by using standard algebraic iterative methods. The solution shows signatures of the coordinate noncommutativity via alterations in the oscillation frequency of the harmonic oscillator system
from its commutative counterpart.  Moreover, it is found that the response of the harmonic oscillator to periodic GW, when their frequencies match, will oscillate with a time scale imposed by the NC parameter. We expect this noncommutative signature to show up as some noise source in the GW detection experiments since the recent phenomenological upper-bounds set on spatial noncommutative parameter implies a length-scale comparable to the length-variations due to the passage of gravitational waves, detectable in the present day GW detectors.} 
%Computation of the 
%expectation value of the particle's position 
%reveals the inherent quantum nature of spacetime noncommutativity.}
\end{abstract}

%\keywords{Noncommutativity, Gravitational Wave}\\

\maketitle
%\noindent {\bf PAC codes:} 11.15.-q, 11.25.-w \\
%{\bf{Keywords:}}Gravitational waves, Noncommutativity\\
%--------------------------------------------
%\section{Introduction}
%--------------------------------------------

Gravitational waves (GW(s)) are tiny vibrations in spacetime 
structure itself. The present day scenario of GW detection 
experiments primarily consists 
of ground-based (LIGO \cite{abramovici}, 
VIRGO \cite{caron}, GEO \cite{luck}, TAMA \cite{ando} etc.) 
and space-based (LISA \cite{lisa}) interferometers. 
The key idea here is to measure the relative optical 
phase shift between the light paths in two perpendicular 
km-length arm cavities where this phase shift is 
due to the relative displacement, 
induced by a passing GW, of the two mirrors 
hung at the end of each arm cavity. 
However, the history of experimental 
GW physics began with resonant-mass detectors, 
pioneered by Weber in the 60's. In the following decades, although 
the sensitivity of resonant-mass detectors have improved considerably, 
it is clear that it could only allow the detection of 
relatively strong GW signals in our Galaxy or in the 
immediate galactic neighbourhood. Nevertheless, 
building the interferometric detectors 
takes many years of preparation and considerable 
finance whereas the resonant detectors, 
being relatively small-scale instruments, 
are more easily realizable. 
Beisdes, the study of resonant-mass detectors 
is instructive in itself because it focuses on 
how a GW interacts with an elastic matter causing vibrations 
with amplitudes many order smaller than the size of a nucleus. 
In a resonant bar it is possible 
to measure vibrations 
which corresponds to just a few tens of phonons \cite{Magg}, 
and variations $\Delta L$ of their length $L$, 
with $\frac{\Delta L}{L} \sim 10^{-19}$. 
It is, therefore, at the quantum mechanical level, 
that experimental evidence for the GWs is likely to appear \cite{Caves}.  
 
Interestingly, in recent developments of noncommutative 
(NC) quantum mechanics \cite{mezin, duv, anisaha, hazra, gov} 
and NC quantum field theory \cite{szabo, doug, masud, bal}, 
where the coordinates $x^{\mu}$ satisfy the NC algebra
\begin{equation}
\left[x^{\mu}, x^{\nu}\right] = i \theta^{\mu \nu}
\label{ncgometry}
\end{equation}
the upperbounds on various NC parameters appearing 
in the literature \cite{carol, bert0, RB, ani, stern, mpr, cst} 
are quite close to this length scale. 
A wide range of theories have been 
constructed in a NC framework including various 
gauge theories \cite{szabo}, gravity \cite{grav} 
and even encompassing certain possible 
phenomenological consequences \cite{jabbari1, rs5, rs7, rs9, rs10, sun}. 
%Naturally, a part of the endeavor is spent in finding the order of various NC parameters and in exploring its connection with observations \cite{mpr, cst, carol, bert0, RB, ani, stern}.
The upperbound on the value of the coordinate 
commutator $\theta^{ij}$ found in \cite{carol} 
is $\lesssim \left(10 {\rm TeV}\right)^{-2}$ 
which corresponds to $4 \times 10^{-40} {\rm m}^{2}$ 
for $\hbar$$=$$c$$=$$1$. 
Whereas such upperbounds on time-space 
NC parameter $\theta^{0i}$ is 
$\lesssim 9.51\times 10^{-18} {\rm m}^{2}$. 
However, recent studies in NC quantum mechanics 
revealed that the NC parameter associated with 
different particles are not same \cite{pmho} and 
this bound could be as high as 
$\theta \lesssim \left(4 {\rm GeV}\right)^{-2} 
- \left(30 {\rm MeV}\right)^{-2}$ \cite{stern}. 
These upperbounds correspond to the length scale 
$\sim 10^{-20} {\rm m} - 10^{-17} {\rm m}$.

 With the prospect of the direct detection 
of GW(s) of such tiny amplitude as $\sim 10^{-18} $ 
in the near future, a good possibility of 
detecting the NC structure of spacetime 
would be in the GW detection experiments 
as it may as well detect the signature of noncommutativity. 
As a first step towards this endeavour, we have analysed the interplay 
of classical GW(s) with a free test particle 
in a {\it NC quantum mechanical} framework in 
\cite{ncgw1}. Our analysis suggested 
that investigating a NC harmonic oscillator system 
interacting with the passing GW is more likely to 
reveal NC effect comparable with the effect of GW. 
Moreover, a simplistic consideration of the interaction 
of GW with the resonant bar shows that the fundamental mode 
of vibration of a thin cylindrical bar is 
formally identical to a harmonic oscillator driven by a 
force exerted by the GW \cite{Magg}. Therefore, 
NC quantum mechanical consideration of the 
interaction of GW with matter, specifically 
the harmonic oscillator which is essentially 
inherent in the resonant mass detector system, 
would certainly prove instructive. 
In light of all these facts, we, in the present paper, 
formulate the quantum mechanics of a NC harmonic oscillator, 
interacting with a linearised GW in the 
long wave-length and low velocity limit. 

Since it has been demonstrated in various 
formulations of NC general relativity \cite{grav}, \cite{grav1}, \cite{banerjee11} 
that any NC correction in the gravity sector is 
second order in the NC parameter, therefore, 
in a first order theory in NC space, 
the linearised GW remains unaltered by 
NC effects and any NC correction appearing 
in the system will be through the matter part only. 
This is true not only with the canonical (i.e. constant) 
form of noncommutativity but also for the 
Lie-algebraic NC spacetime \cite{banerjee11}. 
We shall incorporate the NC effect by writing 
the NC Hamiltonian for the system and then reexpressing 
it in terms of the commutative coordinates and their 
momenta by the well known Bopp-shift transformations 
\cite{stern, cst}. This commutative equivalent 
model will be quantized following $\cite{speli}$. 
We shall deal with the linearly polarized GWs in this analysis for simplicity.  

In a linearized theory of gravity the connection and curvature tensor take the form\footnote{As usual, latin indices run from $1-3$. Also $,$ denotes ordinary derivatives. }  
\begin{eqnarray}
\Gamma^{\mu}{}_{\alpha\beta} &=& \frac{1}{2}\eta^{\mu\nu}\left(h_{\beta\nu, \alpha } + h_{\alpha\nu, \beta } - h_{\alpha\beta, \nu}\right) \label{e1} \\
R_{\alpha\beta\mu\nu} &=& \frac{1}{2}\left(h_{\alpha \nu ,\mu \beta} + h_{\beta \mu,\alpha \nu} - h_{\beta \nu,\alpha \mu } -  h_{\alpha \mu, \beta \nu} \right) \label{e2}
\end{eqnarray}
where $h_{\mu\nu}$ is the metric perturbation on the flat Minkowski background $\eta_{\mu\nu}$. 
Next we choose the 
transverse-traceless (TT) gauge 
\begin{equation}
h_{0\mu} = 0\>,\qquad  h_{\mu\nu;}{}^{\mu} =0\>,\qquad h_\mu^\mu=0
\label{e3}
\end{equation} 
to remove all the gauge redundancies of the theory and the GW is characterised by the only non-zero components \cite{Magg}
%\begin{eqnarray}
$h_{11} = -h_{22}$ and $ h_{12} = h_{21}$, called the $+$ and $\times$ polarisation respectively.
%\label{pol} 
%\end{eqnarray}
The only non-trivial components of the curvature tensor 
in TT-gauge are \footnote{As usual, latin indices run from $1-3$.} 
\begin{equation}
{R^j}_{0,k0} = - \frac{d \Gamma^j_{0k}}{d t} 
= -\frac{1}{2}\frac{d^2 h_{jk}}{d t^2}
\label{e4}
\end{equation}
and the geodesic deviation equation in the 
proper detector frame becomes \cite{Magg} 
\begin{equation}
m\frac{d^2 {x}^{j}}{dt^2} = - m{R^j}_{0,k0} {x}^{k} - m \varpi^{2} x^{j}~.
\label{e5}
\end{equation}
Here $t$ is the coordinate time of the proper detector frame and is same as it's proper time since we are confining ourselves to first order in the metric perturbation.
Eq.(\ref{e5}) governs the response of a 
$2$-dimensional harmonic oscillator with 
frequency $\varpi$ to the passage of a GW. 
Here ${x}^{j}$ is the proper distance of 
the pendulum from the origin, $m$ is its mass
% {\bf{ and $F_{j}$ represents all other forces acting on it.}} 
and the GW is treated as an external classical field. 
Note that, eq.(\ref{e5}) can be used to describe 
the evolution of proper distance in TT-gauge frame 
as long as the spacial velocities involved are 
non-relativistic. Also, $|{x}^{j}|$ has to be 
much smaller than the typical length scale over which 
the gravitational field changes substantially, 
i.e. the reduced wavelength $\frac{\lambda}{2\pi}$ of GW. 
The above conditions are collectively referred 
to as the \textit{small-velocity and long wavelength limit}. 
Thus, with eq.(\ref{e5}) we can analyse the interaction 
of GW with a detector which has a characteristic 
linear size $L \ll \frac{\lambda}{2\pi}$. 
Note that this condition is satisfied by resonant bar 
detectors as well as earth bound interferometers 
but not by the proposed space-borne interferometers 
such as LISA \cite{lisa} or by the Doppler tracking of spacecraft.  

The Lagrangian for the system, whose time 
evolution is described by eq.~$(\ref{e5})$, 
can be written, upto a total derivative term\cite{speli} as 
\begin{equation}
{\cal L} = \frac{1}{2} m\dot {x}^2 - m{\Gamma^j}_{0k}
\dot {x}_{j} {x}^{k}  - \frac{1}{2} m \varpi^{2} x^{2} \>.
\label{e8}
\end{equation}
The canonical momentum corresponding to 
${x}_{j}$ is ${p}_{j} = m\dot {x}_{j} - m \Gamma^j_{0k} {x}^{k}$ 
and hence the Hamitonian becomes
\begin{equation}
{H} = \frac{1}{2m}\left({p}_{j} 
+ m \Gamma^j_{0k} {x}^{k}\right)^2 + \frac{1}{2} m \varpi^{2} x_{j}{}^2  \>.
\label{e9}
\end{equation}
Assuming that the GW is propagating along 
the $z$-axis, due to the transverse nature of 
GW(s), ${\Gamma^j}_{0k}$ has non-zero components 
only in the ${x}-{y}$ plane and therefore the 
response of the pendulum to it is 
essentially confined to that plane. 
To impose noncommutativity on this plane, 
we assume that the coordinates follow 
the algebra (\ref{ncgometry}) and ``quantise" 
the system on this NC plane. 
To proceed, we replace $x^{j}$ and $p_{j}$ in the 
above Hamiltonian by operators ${\hat x}^{j}$ and 
${\hat p}_{j}$ satisfying the 
NC Heisenberg algebra\footnote{We would like 
to mention that it is possible to shift 
the noncommutativity from the coordinates 
to the momenta leading to a dual description as shown in the literature 
\cite{banerjeempla}.} 
\begin{eqnarray}
\left[{\hat x}_{i}, {\hat p}_{j}\right] = i\hbar \delta_{ij} \>, \quad 
\left[{\hat x}_{i}, {\hat x}_{j}\right] = i \theta \epsilon_{ij} \>,\quad 
\left[{\hat p}_{i}, {\hat p}_{j}\right] = 0\>.
\label{e9a}
\end{eqnarray}
It is well known that this can be mapped to the standard 
$\left( \theta = 0 \right)$ Heisenberg algebra 
spanned by $X_{i}$ and $P_{j}$ using \cite{cst, stern}
\begin{eqnarray}
{\hat x}_{i} = X_{i} - \frac{1}{2 \hbar} 
\theta \epsilon_{ij} P_{j}\>, \quad {\hat p}_{i} = P_{i} \>.
\label{e9b}
\end{eqnarray}
Using the traceless property of the GW and 
rewriting the NC version of eq.$(\ref{e9})$ 
in terms of the operators $X_{i}$ and $P_{j}$, we obtain
\begin{eqnarray}
{\hat H} &=& \frac{ P_{j}{}^{2}}{2m} + 
\frac{1}{2} m \varpi^{2} X_{j}{}^{2}+ 
\Gamma^j_{0k} X_{j} P_{k} -
\frac{m \varpi^{2}}{2 \hbar} 
\theta \epsilon_{jm} X^{j} P_{m} \nonumber \\ && 
-\frac{\theta }{2 \hbar} \epsilon_{jm} P_{m} P_{k}  \Gamma^j_{0k} \>. 
\label{e12}
\end{eqnarray}
In the above equation the first two terms are for the ordinary harmonic oscillator, the third term, linear in the affine connections, shows the effect of the passing GW  on the ordinary harmonic oscillator system, the fourth term is the signature of NC space, a pure NC term linear in the NC parameter and the final term shows the coupling between the GW and spatial noncommutativity\footnote{Since we are dealing with linearized gravity, a term quadratic in $\Gamma$ has been neglected in eq.$(\ref{e12})$.}.
% In \cite{carol} the upper-bounds on the coordinate commutator $\theta^{ij}$ have been found to be is $\lesssim \left(10 {\rm TeV}\right)^{-2}$ which corresponds to $4 \times 10^{-40} {\rm m}^{2}$ for $\hbar$$=$$c$$=$$1$. This means a length scale of the order $\lesssim 10^{-20} {\rm m}$. More recent studies in NC quantum mechanics \cite{pmho} revealed that the spatial NC parameter associated with different particles can be different and their upper-bound will be in the range of $\theta \lesssim \left(4 {\rm GeV}\right)^{-2} - \left(30 {\rm MeV}\right)^{-2}$ \cite{stern}, which correspond to the length scale $\sim 10^{-20} {\rm m} - 10^{-17} {\rm m}$. 
Since the length-scale range describing various upper-bounds on NC parameter that we have discussed earlier is quiet close to the detectable variations of length $\Delta L$ in the GW detection experiments, at least the fourth term in the left hand side of equation (\ref{e12}) should be comparable with the third (purely gravitational) term and show up in the GW experiments as some noise source. This, in essence, is similar to some recent results \cite{hogan} where the coordinate noncommutativity is shown to generate holographic noise claimed to be detectable in the signals of the interferometric GW detectors GEO600 \cite{GEO}.

Defining raising and lowering operators 
in terms of the oscillator frequency $\varpi$
\begin{eqnarray}
X_j &=& \left({\hbar\over 2m\varpi}\right)^{1/2}
\left(a_j+a_j^\dagger\right)\>\label{e15a} \\
P_j &=& -i\left({\hbar m\varpi\over 2}\right)^{1/2} 
\left(a_j-a_j^\dagger\right)\>
\label{e15}
\end{eqnarray}
we write the Hamiltonian (\ref{e12}) as 
\begin{eqnarray}
{\hat H} &=& \hbar\varpi\left( a_j^\dagger a_j + 1 \right) 
- \frac{i\hbar}{4} \dot h_{jk} 
\left(a_j a_k - a_j^\dagger a_k^\dagger\right)  \nonumber \\
&& + \frac{m \varpi \theta}{8} \epsilon_{jm} {\dot h}_{jk}  \left(a_{m}a_{k}  - a_{m}a_{k}^\dagger + C.C \right)  \nonumber \\ 
&& -\frac{i}{2} m \varpi ^2 \theta\epsilon_{jk}a_{j}^\dagger a_{k}\>
\label{e16}
\end{eqnarray}
where C.C means complex conjugate. 
Working in the Heisenberg representation, 
the time evolution of $a_{j}(t)$ is given by 
\begin{eqnarray}
\frac{da_{j}(t)}{dt} &=& -i{\varpi}a^j + 
\frac{1}{2}\dot h_{jk}a^\dagger_k - 
\frac{m\varpi^{2}\theta}{2\hbar}\epsilon_{jk} a_{k}\nonumber \\
&& + \frac{i m \varpi \theta}{8 \hbar} \left(\epsilon_{lj} {\dot h}_{lk} 
+ \epsilon_{lk} {\dot h}_{lj}\right)\left(a_{k} - a^\dagger_{k}\right)\>
\label{e17}
\end{eqnarray}
and that of $a_{j}^{\dagger}(t)$ is the C.C of the above equation. 
Next, noting that the raising and lowering operators 
must satisfy the commutation relations
\begin{eqnarray}
\left[a_j(t), a^\dagger_k(t)\right] &=& \delta_{jk}\nonumber \\
\left[a_j(t), a_k(t)\right] &=& 0 = 
\left[a^\dagger_j(t), a^\dagger_k(t)\right]
\label{e18}
\end{eqnarray}
we write them in terms of $a_j(0)$ and $a_j^{\dagger}(0)$, 
the free operators at time $t=0$, 
by the time-dependent Bogoliubov transformations
\begin{eqnarray}
a_j(t) &=& u_{jk}(t) a_k(0) + v_{jk}(t)a^\dagger_k(0)\>
\nonumber \\
a_j^\dagger(t) &=& a_k^\dagger(0)\bar u_{kj}(t)  + a_k(0)\bar v_{kj}(t)\>
\label{e19}
\end{eqnarray}
where the bar denotes the C.C and $u_{jk}$ and $v_{jk}$ 
are the generalised Bogoliubov coefficients. They are
$2\times 2$ complex matrices which, due to eq.~$(\ref{e18})$, must satisfy
%\begin{equation}
$uv^{T}=u^{T}v\>,\> u u^\dagger - v v^\dagger = I,$
%\label{e20}
%\end{equation}
written in matrix form where $T$ denotes transpose, $\dagger$ 
denotes complex conjugate transpose and $I$ is the identity
matrix. Since $a_j(t = 0) = a_j(0)$, $u_{jk}(t)$ and $v_{jk}(t)$ 
have the boundary conditions 
\begin{eqnarray}
u_{jk}(0)& = & I  \quad,\quad  v_{jk}(0) = 0~.
\label{bc0}
\end{eqnarray}
Then, from eq.$(\ref{e17})$ and its C.C, 
we get the following equations of motions in terms of 
$\zeta = u - v^\dagger$ and $\xi = u + v^\dagger$:
\begin{eqnarray}
\frac{d \zeta_{jk}}{dt}=-i\varpi \xi_{jk} 
-\frac{1}{2}{\dot h}_{jl}\zeta_{lk} - 
\frac{m \varpi^{2}\theta}{2 \hbar}\epsilon_{jl} \zeta_{lk}\> 
\label{e21a}\\
\frac{d \xi_{jk}}{dt} = -i\varpi \zeta_{jk} + 
\frac{1}{2}{\dot h}_{jl}\xi_{lk} +
\Theta_{jl} \zeta_{lk} - \frac{m \varpi^{2}\theta}{2 \hbar}
\epsilon_{jl} \xi_{lk}\> 
\label{e21b} 
\end{eqnarray}
where $\Theta_{jl}$ is the term reflecting 
the interplay of noncommutativity with GW
\begin{eqnarray}\Theta_{jl} = 
\frac{i m \varpi \theta}{4 \hbar}\left({\dot h}_{jm}
\epsilon_{ml} - \epsilon_{jm} {\dot h}_{ml}\right)\>.
\label{e21ab}
\end{eqnarray}
Eq(s) $(\ref{e21a}, \ref{e21b})$ are difficult 
to solve analytically for general $h_{jk}$. 
However, our goal, in the present paper, 
is to investigate {\it{whether we get comparable 
effects of spatial noncommutativity and gravitational 
wave on the harmonic oscillator system}} in the 
simplest of settings. 
Therefore we shall solve eq(s) $(\ref{e21a}, \ref{e21b})$ for 
the special case of linearly polarized GW(s). 

\noindent In the two-dimensional plane, the GW, 
which is a $2\times 2$ matrix $h_{jk}$, 
is most conveniently written in terms of the Pauli spin matrices as 
\begin{equation}
h_{jk} \left(t\right) = 2f(t) \left(\varepsilon_{\times}\sigma^1_{jk} 
+ \varepsilon_{+}\sigma^3_{jk}\right) = 2f(t)\varepsilon_A\sigma^A_{jk}\>.
\label{e13}
\end{equation}
Note that the index $A$ runs from $1-3$, 
however, no contribution from $\sigma^2$ 
is included. $2f(t)$ is the amplitude of 
the GW whereas $\varepsilon_{\times} \left(t \right)$ 
and $\varepsilon_{+} \left( t \right)$ 
represent the two possible polarization 
states of the GW and satisfy the condition
$\varepsilon_{\times}^2+\varepsilon_{+}^2 = 1$
for all $t$. In case of linearly polarized 
GW(s) however, the polarization states 
$\varepsilon_{A}$ are independent of time and $f(t)$ is arbitrary. 
To set a suitable boundary condition we shall assume that 
the GW hits the particle at $t=0$ so that 
\begin{equation}
f(t)=0 \>, \quad {\rm for} \ t \le 0.
\label{bc}
\end{equation}
We now move on to solve eq(s) $(\ref{e21a}, \ref{e21b})$ 
by noting that any $2\times 2$ complex matrix 
can be written as a linear combination 
of the Pauli spin matrices and identity matrix. 
Hence we make the ansatz : 
\begin{eqnarray}
\zeta_{jk}\left(t \right) &=& A I_{jk} + B_{1}\sigma^{1}_{jk}  
+ B_{2}\sigma^{2}_{jk} + B_{3}\sigma^{3}_{jk} 
\label{form2}\\
\xi_{jk} \left(t \right) &=&  C I_{jk} + D_{1} \sigma^{1}_{jk} 
+ D_{2} \sigma^{2}_{jk} + D_{3} \sigma^{3}_{jk}~.
\label{form1}  
\end{eqnarray} 
Substituting for $h_{jk}$, $\zeta_{jk}$ and $\xi_{jk}$ 
from eq(s) $(\ref{e13})$, $(\ref{form2})$ and $(\ref{form1})$ 
in eq(s) $(\ref{e21a}, \ref{e21b})$ and comparing the coefficients of 
$I$ and $\sigma$-matrices, we get a set of 
first order differential equations for 
$A, B_{1}, B_{2}, B_{3}, C, D_{1}, D_{2}, D_{3}$ :
\begin{eqnarray}
\dot{A} &=& - i \varpi C - \dot{f}\left(\varepsilon_{1}B_{1} 
+ \varepsilon_{3}B_{3}\right) - i \Lambda B_{2} \nonumber\\
\dot{B}_{1} &=& - i \varpi D_{1} - \dot{f}\left(\varepsilon_{1} A -
 i\varepsilon_{3}B_{2}\right) +  \Lambda B_{3} \nonumber\\
\dot{B}_{2}  &=& - i \varpi D_{2} - i \dot{f}\left(\varepsilon_{3}B_{1} - 
\varepsilon_{1}B_{3}\right) - i \Lambda A \nonumber\\ 
\dot{B}_{3}&=& - i \varpi D_{3} - \dot{f}\left(\varepsilon_{3} A 
+ i \varepsilon_{1}B_{2}\right) - \Lambda B_{1} \nonumber\\ 
\dot{C} &=& - i \varpi A + \dot{f}\left(\varepsilon_{1}D_{1} 
+ \varepsilon_{3}D_{3}\right) + 4 i \lambda \dot{f} 
\left(\varepsilon_{3} B_{1} - \varepsilon_{1} B_{3}\right)\nonumber\\ 
&&-i \Lambda D_{2} \nonumber \\ 
\dot{D}_{1} &=& - i \varpi B_{1} + 
\dot{f}\left(\varepsilon_{1}C - i \varepsilon_{3}D_{2}\right) 
+ 4 \lambda \dot{f} \left(i \varepsilon_{3} A -\varepsilon_{1} B_{2} \right)\nonumber\\ 
&&+ \Lambda D_{3} \nonumber \\ 
\dot{D}_{2} &=& - i \varpi B_{2} + i \dot{f} \left(\varepsilon_{3}D_{1} 
- \varepsilon_{1}D_{3}\right) + 4 \lambda \dot{f} 
\left( \varepsilon_{1} B_{1} + \varepsilon_{3} B_{3} \right) \nonumber\\
&&- i \Lambda C \nonumber \\ 
\dot{D}_{3} &=& - i \varpi  B_{3} + 
\dot{f}\left(\varepsilon_{3} C + i \varepsilon_{1} D_{2}\right) 
- 4 \lambda \dot{f} \left( i\varepsilon_{1} A + \varepsilon_{3} B_{2} \right) \nonumber\\ 
&&-  \Lambda D_{1}  
\label{iteration8} 
\end{eqnarray}
where $\Lambda$ and $\lambda$ are given by 
\begin{eqnarray}
\Lambda = \frac{m \varpi^{2} \theta}{2 \hbar}\quad, 
\quad  \lambda = \frac{m \varpi \theta}{4 \hbar}
\label{parameters}
\end{eqnarray}
and the dot represents derivative with respect to time $t$. 
Noting that $|f(t)|<<1$, the above set of equations can be solved
iteratively about its $f(t)=0$ solution by
applying the appropriate boundary conditions (\ref{bc0}, \ref{bc}).
We therefore obtain to first order in the gravitational wave amplitude and also in the NC parameter 
%\begin{widetext}
\begin{eqnarray}
A(t) & = & C(t)= e^{-i\varpi t}\cos(\Lambda t)
%\left[\cos(\Lambda t) 
%\left(1 + \frac{f^{2}(t)}{2}\right) + i\varpi f(t)\int^{t}_{0}dt^{\prime}
%e^{i\varpi (t - t^{\prime})}\cos(\Lambda t^{\prime})f(t^{\prime})
%\right.\nonumber\\
%&&\left. + \Lambda f(t)\int^{t}_{0} dt^{\prime}e^{i\varpi (t - t^{\prime})}
%\sin(\Lambda t^{\prime})f(t^{\prime}) 
%-\frac{1}{2}\int^{t}_{0} dt^{\prime}e^{i\varpi (t - t^{\prime})}
%(i\varpi \cos(\Lambda t^{\prime})+\Lambda \sin(\Lambda t^{\prime}))
%f^{2}(t^{\prime})\right]
\label{001}\\
B_{2}(t) & = & D_{2}(t) = -ie^{-i\varpi t}\sin(\Lambda t)
%\left[\sin(\Lambda t) 
%\left(1 + \frac{f^{2}(t)}{2}\right) + i\varpi f(t)\int^{t}_{0}dt^{\prime}
%e^{i\varpi (t - t^{\prime})}\sin(\Lambda t^{\prime})f(t^{\prime})
%\right.\nonumber\\
%&&\left. -\Lambda f(t)\int^{t}_{0} dt^{\prime}e^{i\varpi (t - t^{\prime})}
%\cos(\Lambda t^{\prime})f(t^{\prime}) 
%+\frac{1}{2}\int^{t}_{0} dt^{\prime}e^{i\varpi (t - t^{\prime})}
%(-i\varpi \sin(\Lambda t^{\prime})+\Lambda \cos(\Lambda t^{\prime}))
%f^{2}(t^{\prime})\right]
\label{002}\\
B_{1}(t)& = & -D_1(t)= -e^{-i\varpi t}\left[(\varepsilon_{1}\cos(\Lambda t)
- \varepsilon_{3}\sin(\Lambda t))f(t)\right.\nonumber\\
&& \left. + 2i\varpi\varepsilon_1 
\int^{t}_{0}dt^{\prime}
e^{i\varpi (t - t^{\prime})}\cos(\Lambda t^{\prime})f(t^{\prime})
\right.\nonumber\\
&&\left. - 2i\varpi\varepsilon_3
%(i\varpi\varepsilon_3-\Lambda\varepsilon_1)
\int^{t}_{0} dt^{\prime}e^{i\varpi (t - t^{\prime})}
\sin(\Lambda t^{\prime})f(t^{\prime})\right] \nonumber\\
&& + \varpi^{2}\left(\varepsilon_1 \int^{t}_{0} g_1 {(t^{\prime})}dt^{\prime}
-\varepsilon_{3}\int^{t}_{0}g_{2}(t^{\prime})dt^{\prime} \right)
%&&-2i\varpi\Lambda\left(\varepsilon_{3}
%\int^{t}_{0} g_1 {(t^{\prime})}dt^{\prime}
%+\varepsilon_{1}\int^{t}_{0}g_{2}(t^{\prime})dt^{\prime} \right)
\label{003}\\
B_{3}(t) & = & -D_3(t) = -e^{-i\varpi t}\left[(\varepsilon_{3}\cos(\Lambda t)
+\varepsilon_{1}\sin(\Lambda t))f(t)\right.\nonumber\\
&& \left. + 2i\varpi\varepsilon_3 
\int^{t}_{0}dt^{\prime}
e^{i\varpi (t - t^{\prime})}\cos(\Lambda t^{\prime})f(t^{\prime})
\right.\nonumber\\
&&\left. + 2i\varpi\varepsilon_1
%(i\varpi\varepsilon_1+\Lambda\varepsilon_3)
\int^{t}_{0} dt^{\prime}e^{i\varpi (t - t^{\prime})}
\sin(\Lambda t^{\prime})f(t^{\prime})\right]\nonumber\\
&& + \varpi^{2}\left(\varepsilon_3 \int^{t}_{0} g_1 {(t^{\prime})}dt^{\prime}
+\varepsilon_{1}\int^{t}_{0}g_{2}(t^{\prime})dt^{\prime} \right)
%&&+2i\varpi\Lambda\left(\varepsilon_{1}
%\int^{t}_{0} g_1 {(t^{\prime})}dt^{\prime}
%-\varepsilon_{3}\int^{t}_{0}g_{2}(t^{\prime})dt^{\prime}\right)
\label{004}
\end{eqnarray}
%\end{widetext}
where
\begin{eqnarray}
g_{1}(t)=\int^{t}_{0} dt^{\prime}e^{-i\varpi t^{\prime}}
\cos(\Lambda t^{\prime})f(t^{\prime})\nonumber\\
g_{2}(t)=\int^{t}_{0} dt^{\prime}e^{-i\varpi t^{\prime}}
\sin(\Lambda t^{\prime})f(t^{\prime})~.
\label{005}
\end{eqnarray}
The system has now been essentially solved once we specify the initial expectation values of the pendulum's position 
$\vec{r}_{0} = \left(x_{0}, y_{0}\right)$ and momentum $\vec{p}_{0} = \left(p_{x_{0}}, p_{y_{0}}\right)$ 
when the GW just hits the system at time $t=0$. 
The time evolution of the coordinates can be calculated employing the following scheme. Combining the expressions for $A, B_{1}, B_{2}, B_{3}, C, D_{1}, D_{2}, D_{3}$, we can write the solutions for $\zeta$ and $\xi$ using Eq.(s) $(\ref{form2}, \ref{form1})$ which in turn give $u$ and $v$. 
Using Eq.(\ref{e19}), we can now combine $u$ and $v$ into $a_j(t)$ and 
$a_j^\dagger(t)$. From the initial position and momentum expectation values, i.e.$\langle\vec{r}_{0} \rangle= \left(X_{1}(0), X_{2}(0)\right)$ and $\langle\vec{P}_{0}\rangle = \left(P_{1}(0), P_{2}(0)\right)$, we get the raising and lowering operator $a_{j}\left( 0 \right)$ and $a_{j}^{\dagger}\left( 0 \right)$ at time $t = 0$. We then use them in Eq.(s) $(\ref{e19})$ to find $a_{j}\left( t \right)$ and $a_{j}^{\dagger}\left( t \right)$ at a general time $t$ and these yield the time evolution of the expectation values of position coordinates $\langle X_{1} \left(t \right)\rangle$ and $\langle X_{2} \left(t \right)\rangle$ of the pendulum. The general expression of $\langle X_{1}\left(t\right)\rangle$ thus obtained is given by 
\begin{eqnarray}
\langle X_{1}\left(t\right)\rangle & = & [Re(A) + Re(D_{3}^{\star})] X_{1}\left( 0 \right) \nonumber\\
& + & [Im(D_{2}) + Re(D_{1}^{\star})] X_{2}\left( 0 \right)\nonumber\\
& + & [- Im(A) + Im(D_{3}^{\star})]\frac{ P_{1}\left( 0 \right)}{m\varpi} \nonumber\\
& + & [Re(D_{2}) + Im(D_{1}^{\star})]\frac{P_{2}\left( 0 \right)}{m\varpi}
\label{x1}
\end{eqnarray}
where $Re$ and $Im$ denote the real and imaginary parts of the $A$, $D_{1}$, $D_{2}$ and $D_{3}$. Substituting their values from the equations (\ref{001}), (\ref{002}), (\ref{003}) and (\ref{004}) will give the expression for $\langle X_{1}\left(t\right)\rangle$ for a general GW amplitude $f(t)$. 
%Similar expression for $\langle X_{2}\left(t\right)\rangle$ is also obtained. 
Since it will be difficult to see through this much complicated expressions we assume a monochromatic GW waveform oscillating with frequency $\varpi^{\prime}$, $f\left(t\right) = f_{0}e^{i\varpi^{\prime} t}$ which, upon substitution in equation ((\ref{x1}) gives 
\begin{widetext}
\begin{eqnarray}
 \langle X_{1}\left(t\right)\rangle & = & \frac{\left(\cos \varpi_{-} t + \cos \varpi_{+} t\right)}{2} X_{1}(0) + \frac{\left(\sin \varpi_{-} t + \sin \varpi_{+} t\right)}{2 m\varpi}P_{1}(0) + \frac{\left(\cos \varpi_{+} t - \cos \varpi_{-} t\right)}{2} X_{2}(0) + \frac{\left(\sin \varpi_{+} t - \sin \varpi_{-} t\right)}{2m\varpi}P_{2}(0) \nonumber\\
& + & 
\left(1 + \frac{\varpi^{2} \Delta \varpi t}{\Delta \varpi^{2} - \Lambda^{2}}\right)f_{0} \left\{\varepsilon_{3} X_{1}(0) + \varepsilon_{1} X_{2}(0)\right\} - \left(\frac{\varpi^{2} \Lambda t}{\Delta \varpi^{2} - \Lambda^{2}}\right)f_{0} \left\{\varepsilon_{1} X_{1}(0) + \varepsilon_{3} X_{2}(0)\right\} \nonumber\\
& - & 
\left(\varepsilon_{1} + \varepsilon_{3}\right)f_{0} \left[\left(\frac{\varpi^{\prime}_{+}}{\Delta \varpi_{-}}\right)^{2} \sin\frac{\Delta \varpi_{-} t}{2}\left(X_{1}(0)\sin\frac{\Delta \varpi_{-} t}{2} - \frac{P_{1}(0)}{m\varpi}\cos\frac{\Delta \varpi_{-} t}{2}\right) \right.\nonumber\\
&& \qquad \qquad \quad \left. + \left(\frac{\varpi^{\prime}_{-}}{\Delta \varpi_{+}}\right)^{2} \sin\frac{\Delta \varpi_{+} t}{2}\left(X_{2}(0)\sin\frac{\Delta \varpi_{+} t}{2} - \frac{P_{2}(0)}{m\varpi}\cos\frac{\Delta \varpi_{+} t}{2}\right)\right]\nonumber\\
& + & \left(\varepsilon_{1} - \varepsilon_{3}\right)f_{0} 
\left[\left(\frac{\varpi^{\prime}_{-}}{\Delta \varpi_{+}}\right)^{2} \sin\frac{\Delta \varpi_{+} t}{2}\left(X_{1}(0)\sin\frac{\Delta \varpi_{+} t}{2} - \frac{P_{1}(0)}{m\varpi}\cos\frac{\Delta \varpi_{+} t}{2}\right) \right.\nonumber\\
&& \qquad \qquad \quad \left. - \left(\frac{\varpi^{\prime}_{+}}{\Delta \varpi_{-}}\right)^{2} \sin\frac{\Delta \varpi_{-} t}{2}\left(X_{2}(0)\sin\frac{\Delta \varpi_{-} t}{2} - \frac{P_{2}(0)}{m\varpi}\cos\frac{\Delta \varpi_{-} t}{2}\right)\right]
\label{x11}
\end{eqnarray}
where $\Delta\varpi = \varpi - \varpi^{\prime} $, $\varpi_{\pm} = \varpi \pm \Lambda $, $\varpi^{\prime}_{\pm} = \varpi^{\prime} \pm \Lambda $ and $\Delta\varpi_{\pm} = \Delta\varpi \pm \Lambda $. 
\end{widetext}
This result implies that the presence of noncommutativity alters the response of the harmonic oscillator to a periodic GW
from its commutative counterpart.  When the frequency of the GW is very close to that of the harmonic oscillator ($\Delta \varpi \approx 0 $), the oscillatory terms present in the solution representing the response of the system to the GW will oscillate with frequency $\Lambda$ with a large amplitude. It should be possible to detect this effect. 
Now putting $\theta  = 0$, i.e. $\Lambda  = 0$ gives us the classical result of GW interacting with a harmonic oscillator in the low-velocity, long-wavelength limit whereas putting $f_{0} = 0$, i.e. in the absence of gravitational wave the solution assumes the form of a NC harmonic oscillator. 
Interestingly, the presence of $\frac{1}{\hbar}$ factor in $\Lambda$, i.e. in the NC correction terms even after the computation of the expectation value indicates that the NC effect is inherently quantum mechanical in nature. Similar expression for $\langle X_{2}\left(t\right)\rangle$ can also be obtained.  
%From Eq(s) (\ref{x1}, \ref{y1}), 
%we observe that the effect of GWs decreases faster with 
%time for particles with higher momentum. 
%From the above expression for the expectation values, we find that the NC effect in this system increases with time linearly. 
Further realistic scenarios can be obtained if we use various forms of periodic GW signals with more than one frequency and do similar computations. Now that we have studied the interaction of a single NC harmonic oscillator with GW, a further advancement will be to extend it for a macroscopic piece of elastic matter. In fact in a resonant mass detector it is possible to detect vibrations which are incredibly small, with amplitude many orders of magnitude smaller than the size of a nucleus. In that context studying the effect of noncommutativity may prove interesting. Work in this direction is in progress and will be taken up in the subsequent papers. 

%-----------al positio--------------------------
\section*{Acknowledgment} The authors would like to thank the referee for very useful comments.
%-------------------------------------
%\noindent  where a considerable part of this work was 
%completed. The authors would also like to thank the referee for
%useful comments. 

%%%%%%%%%%%%%

%%%%%%%%%%%%%%%%%%%%%%%%%%%%%%%%%%%%%%%%%%%%%%%%%%%%%%%%%%%%%%%%%%%%%%%%%%%%%%%%
\end{document}